\documentclass[traditabstract]{aa} 
\usepackage{graphicx}
\usepackage{txfonts}
\begin{document}
%

\title{The size of AB\,Dor\,A from VLTI/AMBER interferometry
\thanks{Based on observations made with ESO Telescopes at the Paranal Observatory under programme ID 384.C-1053.  }
}

\author{J.C. Guirado\inst{1} \and J.M. Marcaide\inst{1} \and I. Mart\'{\i}-Vidal\inst{2} \and J.-B. Le Bouquin\inst{3}  
\and L.M. Close\inst{4} \and W.D. Cotton\inst{5} \and J. Montalb\'an \inst{6}
}

\institute{ 
Dpto. Astronom\'{\i}a y Astrof\'{\i}sica, Universidad de Valencia, Dr. Moliner 50, 46100 Burjassot, Valencia, Spain\\
\email{guirado@uv.es}
\and 
Max-Planck-Institut f\"ur Radioastronomie, Auf dem H\"ugel 69, D-53121 Bonn, Germany
\and
Laboratoire d'Astrophysique de Grenoble, CNRS-UJF UMR 5571, 414 rue de la Piscine, 38400 Saint Martin d'H\`eres, France
\and
Steward Observatory, University of Arizona, Tucson, Arizona 85721, USA
\and
National Radio Astronomy Observatory, 520 Edgemont Road, Charlottesville, VA 22903-2475, USA
\and
Institut d'Astrophysique et de G\`eophysique de l'Universit\'e de Li\`ege, All\'ee du 6 Ao\^ut 17, 4000 Li\`ege, Belgium
}

\date{Accepted for publication in A\&A}

 
\abstract
{
The pre-main sequence (PMS) star AB\,Dor\,A is the main component of the quadruple system AB\,Doradus.
The precise determination of the mass and photometry of the close companion to AB\,Dor\,A, AB\,Dor\,C,
has provided an important benchmark for calibration of theoretical evolutionary models of low-mass
stars. The limiting factor to the precision of this
calibration is the age of the system, as both the mass and luminosity of AB\,Dor\,A and C are well monitored by other
ongoing programs. In this paper we present VLTI/AMBER observations of AB\,Dor\,A which provide a direct measurement of the
size of this star, 0.96$\pm$0.06\,R$_\odot$. The latter estimate, combined with other fundamental parameters also measured
for this star, allows a precise test of PMS evolutionary models using both H-R diagrams and 
mass-radius relationships. We have
found that our radius measurement is larger than that predicted by the models, which we interpret as an evidence
of the oversizing produced by the strong magnetic activity of AB\,Dor\,A.
Considering, at least partially, this magnetic effect, theoretical isochrones
have been used to derive constraints to the age of AB\,Dor\,A,
favouring an age about 40-50\,Myr for this system. Older ages are not completely excluded by our data.
   }


   \maketitle
%

\section{Introduction}

\noindent
AB\,Doradus is a quadruple stellar system, consisting of two
close pairs, AB\,Dor\,A / AB\,Dor\,C and AB\,Dor\,Ba / AB\,Dor\,Bb, separated by
about 9". The brightest star of the system, AB\,Dor\,A, is a
well-known, pre-main sequence K1-star,
with strong emission at all wavelengths, from radio to X-rays.
Among other instruments, AB\,Doradus has been observed by the
Hipparcos satellite, very-long-baseline-interferometry (VLBI)
arrays (Guirado et al. 1997) and
different near-infrared instruments at the VLT (Close et al. 2005;
Close et al. 2007; Boccaletti et al. 2008). One of the main results from
these observations is the independent
measurement of both the JHK photometry and the dynamical mass
of AB\,Dor\,C (0.090$\pm$0.005 M$_\odot$), the companion to AB\,Dor\,A.
Hence, AB\,Dor\,C is a unique object to calibrate theoretical
mass-luminosity relations; actually, this calibration has
shown that theoretical evolutionary tracks
tend to underestimate the mass of very low mass objects
(Close et al. 2005).

Ongoing observing projects on AB\,Doradus are dedicated to monitoring
both the reflex orbit of AB\,Dor\,A (via VLBI techniques with
Australian antennas) and the differential orbit between AB\,Dor\,A
and AB\,Dor\,C (via near-infrared VLT observations). As the 
above-mentioned observations will improve the photometry and the
dynamical mass of AB\,Dor\,C, the successful calibration of evolutionary models
with AB\,Dor\,C measurements will be limited by the uncertainty in the
estimate of the age of AB\,Dor\,A/AB\,Dor\,C. This parameter is still a
matter of discussion, with different estimates in the literature:
Zuckerman et al. (2004) first proposed an age of 50\,Myr
(later supported by L\'opez-Santiago et al. 2006); on the other hand,
Luhman et al. (2005) and Ortega et al. (2007) estimate an older age for this
system, 120$\pm$20\,Myr; intermediate ages, 75$\pm$25\,Myr, have been reported
by Nielsen et al. (2005), Janson et al. (2007) and  Bocaletti et al. (2008).
This relatively wide range of
ages remains as the largest ambiguity to test model predictions with AB\,Dor\,C.
In this scenario, the measurement of fundamental parameters of any of the members of 
the stellar system may be used to constrain evolutionary models and/or 
derive bounds to the age of the system. In particular, as
PMS stars change in radius as they contract to the main
sequence, having a precise determination of the size of AB\,Dor\,A will
serve, accordingly, to constrain the age of the system.

In this paper we present a precise determination of the size 
of AB\,Dor\,A from VLTI observations performed with the AMBER focal instrument (Petrov et al. 2007), 
installed at the ESO facilities in Cerro Paranal, Chile. In Sect. 2 we describe the observations 
and data reduction; in Sects. 3 and 4 we report the results and discussion, respectively. Conclusions 
are presented in Sect. 5.

\section{Observations and Data Reduction}

\noindent
We observed AB\,Dor\,A with the VLTI using the AMBER instrument at low spectral resolution mode in the J, H, and K bands. 
The observations were performed on 26 December 2009, from 00:30 to 10:30 LST, using the 2m-class Auxiliary Telescopes (ATs) 
placed on stations A0, K0, G1; for cycle 84A, these stations provided the maximum angular resolution for our observation, 
2.3\,milliarcseconds (mas). 
Each AMBER observing block contains 5 exposures of the target (AB\,Dor\,A) or calibrator (see below) and two additional 
exposures for dark and flat correction, each exposure having 200 frames, each recorded with a DIT of 50 milliseconds. 
Target and calibrator were observed alternatively. 
To ensure a proper amplitude calibration, we used three different calibrators, 
namely HD\,35199, HD\,39608, and HD\,39963 (see Table 1), each one with size reported in M\'erand et al. (2006). We selected HD\,35199 as 
primary calibrator and we scheduled it throughout the complete observing run along with AB\,Dor\,A. Not to excessively compromise the 
CAL$-$SCI duty cycle, we included, only after 04:30 LST, and alternatively, observations of HD\,39608 and HD\,39963.
The redundancy provided by multiple calibrators allows a test of the quality of the calibration of the visibility amplitudes.

\begin{table}[t]
\centering
\caption{Parameters of the star calibrators used in our observations.  Values for $\theta_{UD}$ correspond to diameters calculated from
a uniform-disk model (M\'erand et al. 2005).} 
\begin{tabular}{lccc}
\hline
\hline
& Angular distance & K magnitude &  $\theta_{UD}$ \\
& to AB\,Dor\,A ($\degr$) & &  (mas) \\
\hline
HD\,35199  & 2.7 & 3.96 &  0.859$\pm$0.012  \\
HD\,39608  & 5.3 & 3.83 &  0.945$\pm$0.012  \\
HD\,39963  & 2.8 & 4.36 &  0.638$\pm$0.009  \\
\hline 
\end{tabular}
\end{table}

The raw data frames were transformed to individual complex visibilities following the standard routines of the {\sl amdlib} libraries 
(version 2.2; Tatulli et al. 2002) that we outline briefly here. First, using target and calibrators frames, we corrected for the spectral displacement 
between the photometric spectra of the ATs and the interferometric spectrum; second, we removed the bad pixels and applied the 
DARK and FLAT corrections; third, we removed the instrumental dispersive effects, i.e. fringe-fitted each frame using the P2VM 
algorithm in {\sl amdlib}; and fourth, we selected and averaged the frame visibilities resulting from P2VM to obtain a single 
visibility for each exposure and spectral channel. We made extensive tests to determine the appropriate selection criteria (based on 
consistency and robustness of the results; see Sect. 2.1), which were found to be the following: 
we kept frames with atmospheric piston smaller than 
8\,$\mu$m and time between 02:15 and  07:10 (to avoid low-elevation observations), keeping the 50\% of the remaining frames 
with highest SNR (highest fringe contrast).  Finally, we extracted and averaged all the selected exposure visibilities 
within each observing block using python-based algorithms (Mart\'{\i}-Vidal et al. 2011) to obtain the complex visibilities. 
From a first inspection of the visibility amplitudes, we edited out those with low-quality, most of them of the J-band. 
Standard deviations of 2$-$5\% were obtained for the remaining visibility amplitudes.

\begin{figure}[t]
\vspace{9cm}
\includegraphics{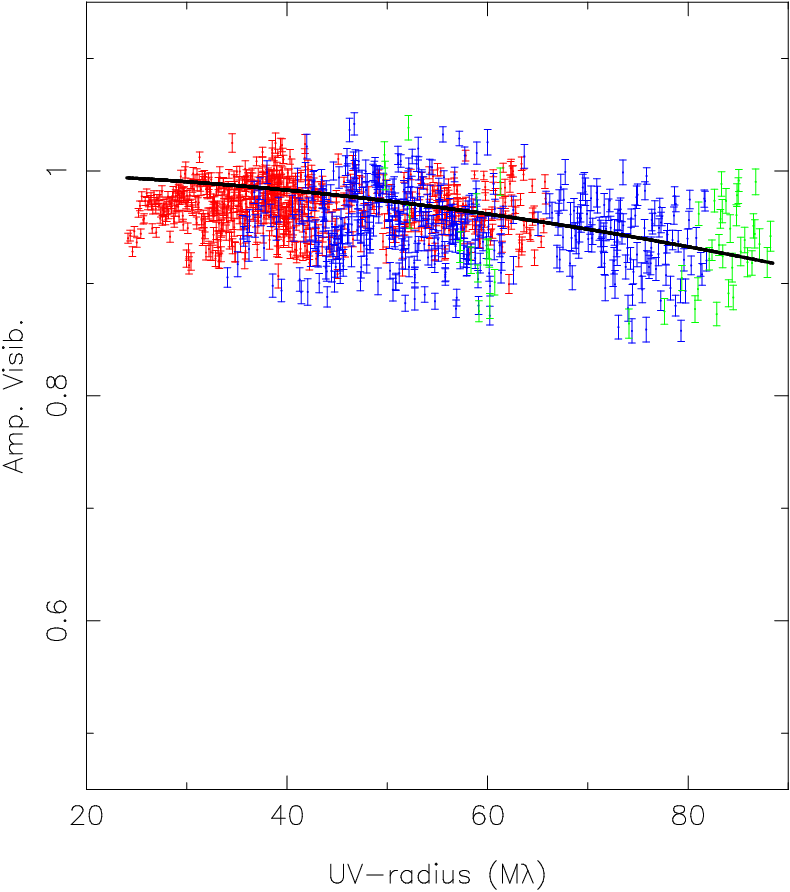}
\caption[] 
{\small AMBER/VLTI squared visibility amplitudes of AB\,Dor\,A for J, H, and K bands (green, blue, and red data, respectively). The black continuous 
line corresponds to the visibilities resulting from the best fit of the data to a uniform disk model. The source is only partially resolved 
by the triplet A0-K0-G1, but sufficiently to constrain the angular size of AB\,Dor\,A to 0.62$\pm$0.04\,mas.}
\end{figure}

\subsection{Amplitude visibility and wavelength calibration}

\noindent
AMBER amplitude is usually calibrated by measuring the transfer function of a star with known size, which is 
interpolated to the target star times, assuming the transfer function remains unchanged (Tatulli et al. 2002). In our observations, the transfer 
function was calculated by comparing the measured visibility amplitudes of the main calibrator HD\,35199 with the predictions from a 
uniform-disk (UD) model with the diameter shown in Table 1. The accuracy of the estimate of the size of AB\,Dor\,A depends to a high degree on 
the quality (i.e. spatial and temporal stability) of the transfer function obtained from HD\,35199.  To ascertain this quality, 
we used HD\,35199 to calibrate the amplitudes of the other two CAL stars, HD\,36908 and 
HD\,39963, both with known size as displayed in Table 1. Should the HD\,35199 be well edited and calibrated, a modelfit process must provide 
UD-diameters for the two secondary calibrators similar to those in Table 1. We found optimal coincidence between 
tabulated and modelfitted diameters for the selection criteria described in Section 2.  

The lack of spectral calibration for AMBER leads to an uncertainty in the observing wavelength $\lambda$. This 
effect can be manually corrected by means of the identification of absorption lines present in the data of the target 
star and/or the calibrators (i.e. Mart\'{\i}-Vidal et al. 2011). However, finding absorption features in LR-mode AMBER observations  
is difficult; in fact, our data does not show any atmospheric signature pronounced enough to be used for a precise spectral
calibration. Hence, we were forced to adopt a different approach, using again the two secondary calibrators in a similar procedure 
to that described above.  We explored the solution (i.e. UD-diameter for both secondary calibrators) using shifts in $\lambda$ between $-$0.15 and 
0.15\,$\mu$m; this quantity looks conservative in view of other estimates at low- and medium-resolution AMBER observations 
(Domiciano de Souza et al. 2008;  Krauss et al. 2009; Mart\'{\i}-Vidal et al. 2011).  Within this range, we found that a shift of $+$0.1\,$\mu$m minimized the 
differences between our modelfitted sizes and the nominal sizes displayed in Table 1.  
Final modelfit estimates of the UD-diameters were 
0.95$\pm$0.04\,mas for HD\,39608 and 
0.66$\pm$0.04\,mas for HD\,39963, 
where the standard deviations shown included the contribution of this wavelength calibration process 
(we conservatively increased the statistical errors, $\pm$0.02\,mas, to cover the scatter found in our exploratory approach.) Once the 
goodness of the transfer function was assessed, we calibrated the visibility amplitudes of AB\,Dor\,A similarly, 
using HD\,35199 data.

%
\begin{figure}[t]
\vspace{13cm}
\includegraphics{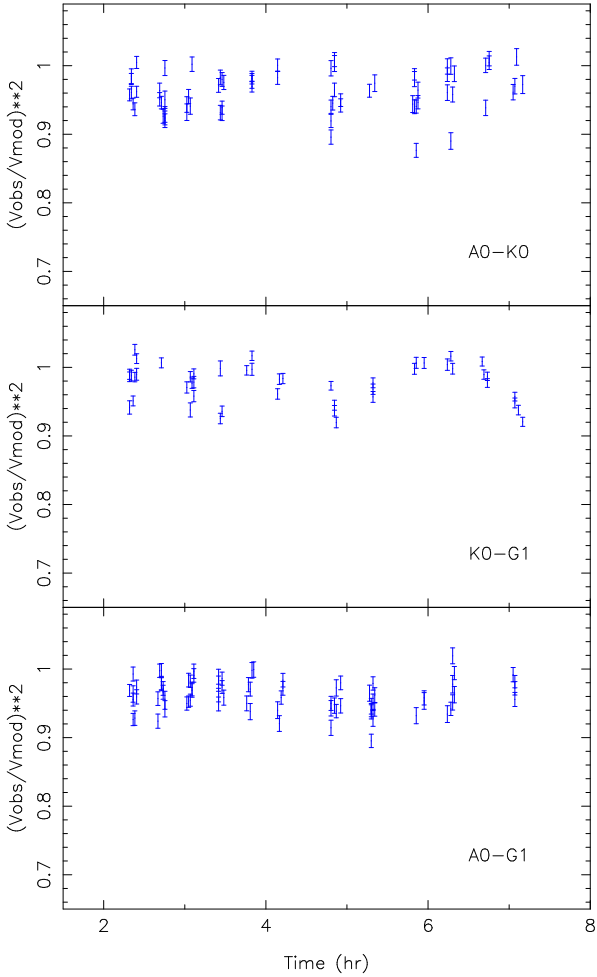}
\caption[] 
{\small Time evolution of the ratio between observed and modeled amplitude visibilities 
for the triplet A0-K0-G1 (H-band). Data have been 
averaged in 0.5\,M$\lambda$ bins in the {\it uv}-plane. These residuals uncover the substructures not 
accounted for by the uniform disc model. See text.} 
\end{figure}
%

\section{Results}

\noindent
The observed and calibrated visibilities of AB\,Dor\,A (see Fig. 1) show that AB\,Dor\,A is partially resolved by the A0-K0-G1 interferometer, as 
expected for a solar-size star placed at the AB\,Dor\,A distance. Modelfit of the visibility amplitudes to a uniform-disk model gives a size of AB\,Dor\,A of 
0.62$\pm$0.04\,mas. 
We note that the UD model is a crude approximation of the stellar surface brightness distribution, and that other models may 
provide a more realistic representations, i.e., a limb-darkened disk model (LD model), with decreasing intensity toward the edge 
of the stellar disk. The estimate of a LD diameter directly from the AB\,Dor\,A visibilities looks difficult, given the limited sampling of 
the first lobe of the visibilities, which is probably not enough to discriminate between different LD sizes.
Instead, following Di Folco et al. (2004), we converted the UD-diameter ($\theta_{UD}$) to LD-diameter ($\theta_{LD}$) using the approximate 
expression given in Hanbury-Brown et al. (1974).\\  

\begin{equation}
\rho(\lambda) = \frac{\theta_{UD}}{\theta_{LD} }\sqrt{ \frac{1-u(\lambda)/3 }{1-7u(\lambda)/15 }  }
\end{equation}

\noindent
where the coefficients $u(\lambda)$ for our spectral bands were taken from the tables in Claret (2000). The use of Eq. 1 with the K-band coefficient 
provides a conversion factor $\rho$ of 1.027 which, in turn, yields a LD diameter for AB\,Dor\,A of 0.60$\pm$0.04\,mas. Given the small magnitude of 
this LD correction (less than 3\%), the uncertainties assigned to the coefficient $u(\lambda)$ (up to 10\%) do not significantly alter the radius estimate. 
Similarly, the use of the H-band coefficients 
in Eq. 1 does not affect the LD-diameter by more of 0.3\%.  
Finally, using the distance measurement to this system, 14.9$\pm$0.1\,pc (Guirado et al. 2006), 
the LD angular diameter can be readily converted into a linear diameter of AB\,Dor\,A: 0.96$\pm$0.06\,R$_\odot$ (see Table 2). 
Given the extraordinary precision of the distance determination, we note that the uncertainty in the 
latter angular-to-linear conversion is just that corresponding to error propagation. 

\begin{table}[t]
\centering
\caption{Fundamental parameters of the PMS star AB\,Dor\,A. }
\begin{tabular}{lc}
\hline
\hline 
K magnitude: & 4.686$\pm$0.016\tablefootmark{a} \\         
Luminosity (L$_\odot$): & 0.388$\pm$0.008\tablefootmark{a} \\         
T$_{eff}$ (K) & 5081$\pm$50\tablefootmark{a} \\         
$v$\,sin\,$i$ (km/s) & 91$\pm$1\tablefootmark{b}\\ 
Distance (pc): & 14.9$\pm$0.1\tablefootmark{c} \\         
Mass (M$_\odot$): &  0.86$\pm$0.05\tablefootmark{d} \\
$\theta_{UD}$ (mas): & 0.62$\pm$0.04\tablefootmark{e} \\
$\theta_{LD}$ (mas): & 0.64$\pm$0.04\tablefootmark{e} \\
Radius (R$_\odot$): & 0.96$\pm$0.06\tablefootmark{e} \\
\hline
\end{tabular}
\tablefoot{
\tablefoottext{a}{Close et al. (2007).}
\tablefoottext{b}{Collier-Cameron et al. (2002).}
\tablefoottext{c}{Guirado et al. (1997).}
\tablefoottext{d}{Guirado et al. (2006).}
\tablefoottext{e}{This paper.}
}
\end{table}

\begin{figure}[t]
\vspace{19cm}
\includegraphics{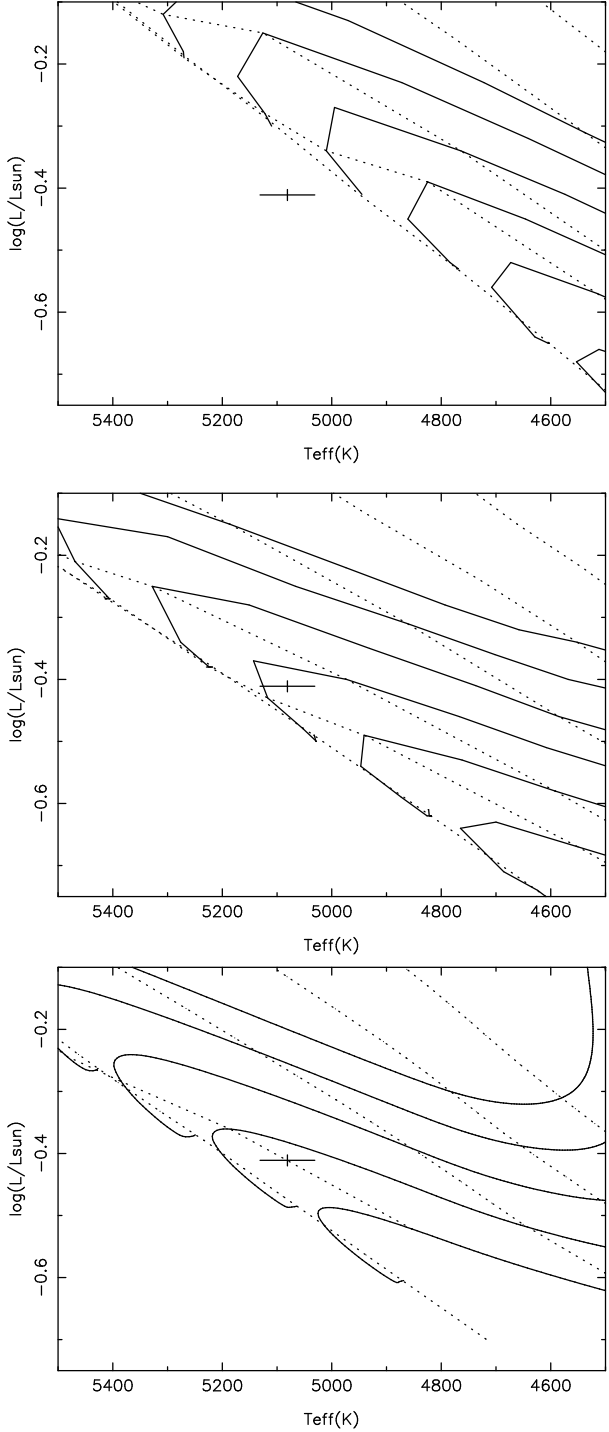}
\caption[] 
{\small H-R diagrams for several PMS evolutionary models. Isochrones (dotted lines) correspond to 10 (top isochrone), 16, 25, 40, and 100 Myr (an additional 
50\,Myr isochrone is shown 
in the middle plot). 
Isomasses (continuous lines) 
are for 0.75 (bottom isomass), 0.80, 0.85, 0.90, 0.95, and 
1.0 M$_{\odot}$. Points with error bars correspond to measurements (see Table 2). {\it (Top)} BCAH98 models, $\alpha=1$. 
{\it (Middle)} BCAH98 models, $\alpha=1.9$ (solar value). 
{\it (Bottom)} MDKH04 models.}
\end{figure}

\begin{figure}[t]
\vspace{19cm}
\includegraphics{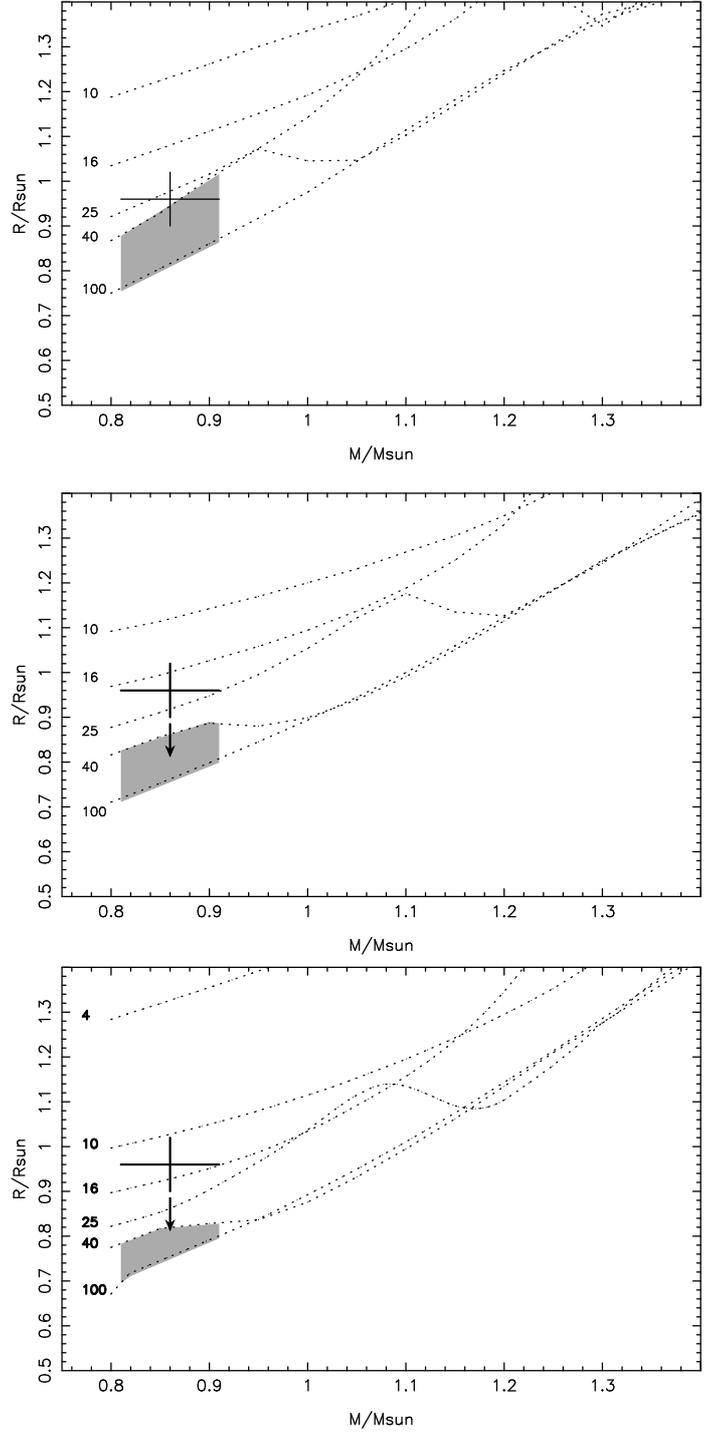}
\caption[] 
{\small Mass-Radius relationship for several PMS evolutionary models. Isochrones (in Myr) are shown as dotted lines. Points with error bars correspond to measurements. 
The shadowed area indicates the age range considered for the low-mass companion AB\,Dor\,C (40$-$120\,Myr). 
Arrows in middle and bottom plots indicate the "magnetic free" radius (see text).
{\it (Top)} BCAH98 models, $\alpha=1$. 
{\it (Middle)} BCAH98 models, $\alpha=1.9$ (solar value). 
{\it (Bottom)} MDKH04 models.}
\end{figure}

\subsection{Sub-structures in the visibility amplitude}

\noindent
Despite the scatter of the measurements, the visibility amplitudes in Fig. 1 show small sinusoidal trends, up to 10\%,
for K and H bands. Clearly, these substructures are not accounted for by 
our uniform disk model. A natural explanation could be the presence of the low-mass companion AB\,Dor\,C. However, this is not likely: 
the separation between AB\,Dor\,A and AB\,Dor\,C at the time of the VLTI observations is $\sim$300\,mas (calculated with the orbital parameters 
given in Guirado et al. 2006), which places AB\,Dor\,C out of the field-of-view of the AMBER instrument when used with the ATs (250\,mas). Therefore, 
amplitude variations due to the small companion are not expected for such a particular position in the orbit to within the AMBER capabilities.

Rather, we consider that such substructures may be due to stellar activity on the surface on AB\,Dor\,A, most probably a combination 
of starspots and stellar rotation. The rotation and magnetic activity of the AB\,Dor\,A surface has been well studied
(i.e. Collier-Cameron \& Donati 2002; Cohen et al. 2010 and references therein), and the presence and evolution 
of starspots characterized. Typically, spots permanently cover a significant portion of the stellar surface; to complicate the scenario, the 
spots move quickly over the surface due to the fast rotation of AB\,Dor\,A. 
Since our observation time span is comparable to the rotation period of AB\,Dor\,A (0.514 days; Innis et al. 1998) it is not a surprise that the visibilities could be 
affected by such fast structural variations. Actually, this effect is better seen in Fig. 2, which shows the time evolution of the residuals of 
the amplitude visibilities, expressed 
as the ratio between the observed and modeled values, for the triplet A0-K0-G1 (for clarity, data have been 
averaged in 0.5\,M$\lambda$ bins in the {\it uv}-plane). These residuals uncover the sinusoidal signature present in the amplitude 
visibilities that we assigned to the presence of starspots in fast rotation.
We have carried out simple interferometric tests to verify the above statement: the starspots can be simulated by the addition of smaller 
disks on the surface of our uniform disk; such a smaller disks introduce an asymmetry in the AB\,Dor\,A structure, which, as seen by an interferometer, 
effectively produce a sinusoidal signature in the visibility amplitudes. However, more precise observations and a more elaborated 
model of the rotation, size, variability, and number of starspots in AB\,Dor\,A, both beyond the scope of this paper, appear necessary to properly 
clarify the origin of the subtle variations in our interferometric observables. 

\section{Discussion}

\subsection{Comparison with PMS stellar models}

\noindent
Our radius measurement of AB\,Dor\,A is in reasonable agreement with previous estimates based on other techniques, i.e. techniques based on the  projected equatorial velocity 
($\sim$1\,R$_\odot$; Maggio et al. 2000), and techniques based on the empirical Barnes et al. (1978) relationship between radius and $V-R$ colors 
(0.98$\pm$0.04\,R$_\odot$; Collier-Cameron \& Foing 1997; Maggio et al. 2000). 
All these results indicate, as already pointed out by Collier-Cameron \& Foing (1997), 
that AB\,Dor\,A has not yet reached the main sequence, as its radius is larger than that for a ZAMS star of the same spectral type. 
AB\,Dor\,A is one of the few PMS stars having a very precise determination of many of its fundamental parameters (see Table 2); hence, they can be compared with 
the values resulting from PMS evolutionary models, both in the luminosity/effective temperature plane (H-R diagram) and in the mass/radius plane (M-R plane). 
Eventually, we should derive an age for the system both from the H-R diagram and for the M-R plane. We have used the PMS models from Baraffe et al. (1998; BCAH98) 
and Montalb\'an et al. (2004; MDKH04). The BCAH98 models use the NextGen atmospheres (Hauschildt et al. 1999) with a treatment of the convection based on the 
mixing length theory (MLT; B\"ohm-Vitense 1958). MLT convection is characterized by the mixing parameter $\alpha$, defined as $l_{mix}/H_p$, with $l_{mix}$ 
the convective mixing length and $H_p$ the pressure scale height; in particular, we used models with $\alpha=1$ and $\alpha=1.9$, the latter being required to 
match the solar values. In contrast, the MDKH04 models use the atmospheres from Heiter et al. (2002)  with a convection treatment based on full spectrum 
turbulence (FST; Canuto \& Mazzitelli 1991; Canuto et al. 1996). We used FST models made to fit the solar values; this makes MDKH04 models more comparable 
to BCAH98 with $\alpha=1.9$.

In Fig. 3 we show the H-R diagram for the three models considered, where the placement of AB\,Dor\,A can be compared with 
theoretical isochrones and isomasses. As seen in the Figure, BCAH98 models for $\alpha=1$ fail to predict any
of the measurements of AB\,Dor\,A. However, both BCAH98 with $\alpha=1.9$ and MDKH04 models offer 
good predictions for both age and mass. Regarding the mass, both models predict a mass for AB\,Dor\,A below the dynamical mass estimate 
(0.86$\pm$0.05\,M$_\odot$), but well within the quoted uncertainties. The slight underprediction displayed in these two models is in agreement with  
that reported in Hillenbrand \& White (2004) and Mathieu et al. (2007), who pointed out that PMS models may underpredict masses 
below 1.2\,M$_\odot$. 
On the other hand, isochrones in H-R diagrams of BCAH98 with $\alpha=1.9$ and MDKH04
seem to favor an early age for AB\,Dor\,A (a range of 40-50\,Myr covers the estimates of both models). However, 
as we will see below, success (or failure) in the H-R diagram may not translate to the M-R plane.

Comparisons in the M-R plane could be considered even more fundamental than those in the H-R diagram, as they depend
only of the accuracy of the measurements, free from ambiguities related to the determination of effective temperatures 
(Mathieu et al. 2007). In Fig. 4, our measurement of both mass and radius for AB\,Dor\,A are shown along with isochrones corresponding to the same models as in Fig. 3. 
The shadowed area corresponds to a generous range of possible ages for AB\,Dor\,C, the low-mass companion of 
AB\,Dor\,A (40-120\,Myr; see Sect. 1). In contrast to the results obtained above in the H-R diagram, BCAH98 models with $\alpha=1$ seem 
to agree, within uncertainties, with the younger side of the AB\,Dor\,C age interval, while BCAH98 with $\alpha=1.9$ 
and MDKH04 models predict an age for AB\,Dor\,A which is perhaps too young (16-25\,Myr and 10-16\,Myr, respectively) 
if we posit that AB\,Dor\,A and AB\,Dor\,C are coeval. This apparent mismatch between the predictions of the H-R diagram and the M-R plane has been 
already reported for other stars (i.e. Stassun et al. 2004). In the case of AB\,Dor\,A, the strong magnetic activity of this star 
may play an important role in the predictions of both the H-R diagram and the M-R plane. We discuss this in turn.

\subsection{The role of the magnetic field}

\noindent
There is some evidence that the magnetic activity may influence the evolution and structure of PMS objects. Torres et al. (2006) found some disagreement
($\sim$10-15\%) between predicted and modeled radius for PMS active eclipsing binaries which was not for non-active stars. The connection between magnetic activity 
and size has been reported earlier (Ribas 2003; Torres et al. 2006 and references therein): in essence, the stellar convective heat transport 
is inhibited by the presence of a strong magnetic 
field, usually related to a rapid stellar rotation, and by the frequency and duration of spots in the stellar surface. As a consequence of this 
loss of efficiency of the convection, the star must augment its size to radiate the accumulated energy; therefore, magnetically active stars 
would have larger radii than those estimated in absence of magnetic field. Following the same line of reasoning, in the framework of the mixing length treatment
of convection, a less efficient convection corresponds to a lower value of the mixing parameter $\alpha$ (Tayler 1987). In fact, Torres et al. (2006) also found 
that models with lower $\alpha$ agree better with the estimated size of active stars. 

AB\,Dor\,A is a highly-magnetized (surface values of 200\,G on average; Cohen et al. 2010), 
fast-rotating star and certainly a good candidate 
to suffer the effects of a strong magnetic field on its evolution.  Actually, the BCAH98 isochrones for $\alpha=1$ are compatible at the same 
time with the measured radius of AB\,Dor\,A and, to within uncertainties, 
with the age interval of AB\,Dor\,C (see Fig. 4). On the other hand, the radius of AB\,Dor\,A predicted by the BCAH98 with $\alpha=1.9$ and MDKH04 in 
the age interval of AB\,Dor\,C is smaller than that measured. Hence, assuming coevality, the available data indicate that convection in AB\,Dor\,A 
must be less efficient than in the Sun, and that the required value of the mixing length parameter to fit the radius of AB\,Dor\,A is smaller than that 
required to fit the radius of the Sun at its present age.  Based on the results of Torres et al. (2006), 
we can estimate a "magnetic-free" radius of AB\,Dor\,A allowing for an oversizing factor of 15\% in our 
interferometric measurements as a consequence of the strong magnetic activity. If we translate this "magnetic-free" radius ($\sim$0.81\,R$_{\odot}$) to the 
M-R plane representations (indicated by arrows in Fig. 4, middle and bottom plots), the measurements would be placed well in the range of the AB\,Dor\,C age, 
actually in agreement with the age predicted by the same models from the H-R diagrams (40-50\,Myr). 

This agreement between the age estimates from H-R diagrams and M-R planes seems to favour 
the younger side of the AB\,Dor\,C range as the most probable age for the AB\,Doradus system, substantially younger than that of the Pleiades 
cluster ($\sim$120\,Myr; Luhman et al. 2005). Our result is marginally compatible with the latter, older age, just at the extreme end of the uncertainties associated 
with our radius measurement.\\ 

It should be mentioned that the fast rotation of AB\,Dor\,A has further consequences beyond those related with the magnetic field. The most direct effect is that 
the high rotation may alter the 
oblateness of the star producing a larger radius at the equator than at the poles. In turn, the oblateness the star leads to gravitational darkening 
(von Zeipel 1924); according to this effect, both the surface gravity and the brightness of the star decrease from the poles to the equator. 
Both effects could be present in AB\,Dor\,A since, in fact, there are evidences of a time-dependent oblateness in this object (Collier-Cameron \& Donati 2002). 
On the other hand, oblateness and gravitational darkening have been measured from IR interferometry for other stars like Achernar or Altair (Domiciano de Souza et al. 2003; 2005). 
However, the above-mentioned stars subtend an angular size $\sim$5 times larger than that of AB\,Dor\,A (0.62$\pm$0.04\,mas), which is only partially resolved by our AMBER/VLTI 
observations (see Fig. 1). Therefore, the detection of these fine details of the structure of AB\,Dor\,A would require a longer and more sensitive interferometer that should shed 
some light on these important, rotationally-related, contributions.\\

Finally,  our angular size estimate of AB\,Dor\,A can be combined with the bolometric flux to obtain a direct measurement of the effective temperature from the Stephan-Boltzmann law. 
For this purpose we have used a bolometric correction at K-band, BC(K)=1.9$\pm$0.1, obtained from the polynomial fits of Masana et al. (2008). 
The resulting value of the effective temperature is 4800$\pm$300\,K, coincident, within uncertainties, to the temperature reported in Table 2. We note that the large 
uncertainties of this effective temperature mostly correspond to the $\sim$6\% error in the radius estimate. 

\subsection{Influence of the metallicity}

\noindent
G\'omez de Castro 2002 reported a metallicity for AB\,Dor\,A of [M/H]=$-$0.3. We have used the MDKH04 models to evaluate the effect of the metallicity in the 
M-R plane. In Fig. 5 we show the isochrones for both solar composition and [M/H]=-0.3. The age inferred from our measurement is not altered significantly following
one or other value of the metallicity.  As done in Fig. 4, an arrow accounts for the 15\% oversizing, which points to ages younger than 40\,Myr. However, 
we notice that, for [M/H]=$-$0.3, isochrones are packed between 40 and 100\,Myr, making it difficult to choose a 
particular age within that range.  

\noindent
\begin{figure}[t]
\vspace{6.5cm}
\includegraphics{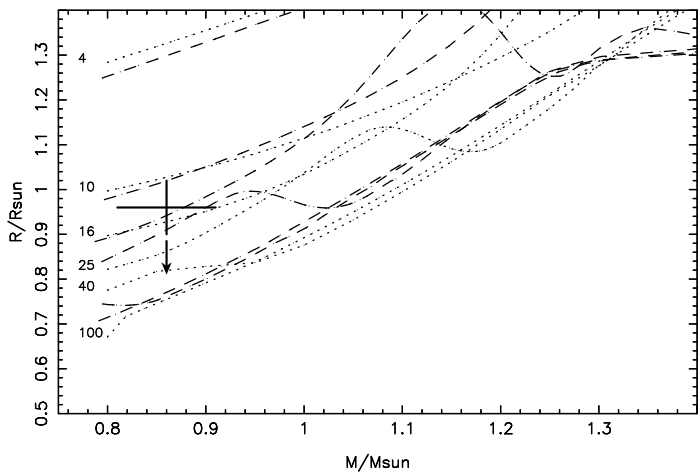}
\caption[] 
{\small Mass-Radius relationship for different metallicities (MDKH04 models). Isochrones (in Myr) are shown for [M/H]=0 (dotted lines) and 
[M/H]=$-$0.3 (dashed line, which corresponds to the AB\,Dor\,A metallicity reported in G\'omez de Castro 2002). }
\end{figure}

\section{Conclusions} 

\noindent
We present new AMBER/VLTI interferometry observations determining the size of the PMS star AB\,Dor\,A. We have used a simple model 
(uniform disk) to fit the interferometric visibilities and derive an angular diameter of 0.62$\pm$0.04\,mas. The corresponding limb-darkened value 
(0.64$\pm$0.04\,mas), combined with a very precise previous estimate (14.9$\pm$0.01\,pc), allows for a most precise measurement of the linear radius 
(0.96$\pm$0.06\,R$_\odot$). 
Some (weak) substructures are also apparent in the visibilities (sinusoidal variations up to 10\% in amplitude). 
We consider that these sinusoidal variations might be produced by stellar spots, which are frequent and intense in AB\,Dor\,A as a consequence of its chromospheric 
activity. The fast rotation rate of AB\,Dor\,A, comparable to our observation time span, may play a role too, masking the visibility trends. 
The combination of dynamical mass and this new radius determination facilitates the comparison of these two fundamental parameters with those 
provided by theoretical PMS stellar models. We have found evidence of disagreement between the predictions based on H-R diagrams and those based 
on M-R planes. Part of this discrepancy could be due to the strong magnetic field on the surface of AB\,Dor\,A, which, as other authors 
point out and our results show, may inhibit the efficiency of the convection and produce a larger radius than predicted by PMS models calibrated to fit the 
radius of the Sun.  Should this magnetic effect be accounted for, we could reconcile the predictions from the H-R diagram and the M-R plane for the models 
considered in this paper (except the H-R predictions of BCAH98 models with $\alpha=1$), favouring an age for AB\,Dor\,A of 40-50\,Myr, 
at the younger end of the range of published ages of the low-mass companion AB\,Dor\,C. Older ages are 
not completely excluded by our work, although coevality with the Pleiades cluster appears to be marginal and only compatible with our data at the extreme end of 
the (somewhat conservative) size uncertainties.  Finally, we notice that, with this new estimate of the linear radius, AB\,Dor\,A is one of the few PMS stars 
with most of the fundamental parameters precisely determined (see Table 2). This makes AB\,Dor\,A a very appropriate object to check the consistency of PMS 
models, and in particular, those intended to include the effects of the magnetic activity in stellar evolution. 

\begin{acknowledgement}
This work has been partially founded by grant AYA2009-13036-C02-02 of the Spanish MICINN, and by 
grant PROMETEO 104/2009 of the Generalitat Valenciana. The National Radio Astronomy Observatory is operated by Associated Universities, Inc. 
under cooperative agreement with the (U.S.) National Science Foundation.
This research has made use of the SIMBAD database, operated at CDS, Strasbourg, France.
\end{acknowledgement}

\end{document}